\documentclass[twocolumn,showpacs,preprintnumbers,amsmath,amssymb]{revtex4}

\usepackage{graphicx}
\usepackage{dcolumn}
\usepackage{bm}

\begin{document}

\title{Quantum physical relevance of the Einstein tensor}

\author{J. Lamey}
\email{jakob.lamey@physik.uni-regensburg.de}
\author{G.M. Obermair}%

\affiliation{Department of Physics, University of Regensburg, D-93040 Regensburg, Germany}

\date{\today}

\begin{abstract}
Taking quantum physics as well as large scale astronomical observations into account, a spacetime metric is introduced, such that the nonlinear part of the Einstein tensor contains effects of the order\,\(\hbar\).
\end{abstract}

\maketitle

\section{Introduction}

On large scales general relativity unveils that gravity appears as
curvature of spacetime. Gravity and spacetime exist everywhere.
It should therefore to be expected that doing quantum physics, learning
more and more about the behavior of nature on small scales, would
create an understanding of quantum physics in terms of curvature not only qualitatively but also quantitatively.

But this has not been the case so far in spite of the fact that the
very notion of curvature itself, in the sense of obstruction to flatness,
contains information on the internal structure of a curved manifold which in turn influences the local dynamics.

Nevertheless, with the Casimir and the Aharonov-Bohm effect two experimental
observations are given in the quantum physical realm, which may give
hints for an understanding of the role of curvature in quantum physics.
As yet these experiments could not be explained in the context
of general relativity. 
There are, however, new large scale observations from astronomy
requiring analysis both along the lines of the Casimir and
the Aharonov-Bohm effect as well as those of general relativity.

This text is organised as follows. After a short account of the Casimir
and the Aharonov-Bohm effect, sec.II, we introduce two large scale observations: galaxy superclusters and the time development of the scale factor. Both are discussed with respect to their impact on quantum physics and general relativity, sec.III.
In sec.IV we define values and formulas which we use in sec.V to introduce a spacetime metric such that its curvature properties yield quantum physical orders of magnitudes. Consequences regarding the physical meaning of gravitational potentials in spacetime and the specific nonlinearity of general relativity are discussed. 

The following convention is used. The metric tensor \( g_{\mu \nu } \)
is taken to have Lorentzian signature of type (-,+,+,+), the Ricci
tensor \( R_{\mu \nu } \) is obtained from the Riemannian curvature
tensor \( R_{\mu \nu \rho \sigma } \) by contraction over the first
and the fourth index, \( R_{\mu \nu }=R^{\rho}_{\mu \nu \rho} \).

The Einstein field equations in geometric units are 
\begin{equation}
\label{Einsteinefeq}
G^{\mu \nu }=8\pi T^{\mu \nu }.
\end{equation}
Here \(G^{\mu \nu }\) is the Einstein tensor,
\begin{equation}
\label{Einsteintensor}
G^{\mu \nu }=R^{\mu \nu }-\frac{1}{2}Rg^{\mu \nu }, 
\end{equation} 
where \(R\) is the scalar curvature and \(T^{\mu \nu }\) the energy momentum tensor.

\section{On Casimir and Aharonov-Bohm effect}

Calculating quantum fluctuations of the electromagnetic field between two perfectly conducting plates, placed parallel in vacuum, Casimir \cite{Casimir} predicted an attractive force \( F \) per unit area \( A \) between these two plates at a mutual distance \(d\)
\begin{equation}
\label{Casimirgew}
\frac{F}{A}=-\frac{\pi ^{2}\hbar c}{240d^{4}}.
\end{equation}
  For a review of Casimir effect see monograph \cite{Mostepanenko}.

Remarkably, no constant characterizing an interaction is appearing
in formula (\ref{Casimirgew}). This suggests the Casimir force
to be of geometrical origin. In addition, the Casimir
force per unit area is in fact a pressure. Due to general relativity
and the Einstein field equations (\ref{Einsteinefeq}), any kind of pressure contributes to the curvature of spacetime. Thus an expression
of Casimir force per unit area in terms of curvature should be expected.

Indeed, in geometric units \( c \) and the gravitational
constant \( G \) are put equal to one and Planck's constant \( \hbar  \) becomes about \( \hbar^{gu}=2.6\times 10^{-66}{\rm cm}^{2} \)  \cite{Waldgrbook}. Writing (\ref{Casimirgew}) for
unit area \(A=1{\rm cm}^{2}\) and a distance \(d=1{\rm cm}\) we get
\begin{equation}
\label{Casimirgu}
\frac{F}{A}=-0.04\frac{\hbar^{gu }}{{\rm cm}^{4}}=-0.04\frac{2.6\times 10^{-66}}{{\rm cm}^{2}}.
\end{equation}
 This has the physical dimensions of intrinsic curvatures, like Gaussian,
scalar or sectional curvatures for example.

Aharonov and Bohm \cite{Aharonov Bohm orig} theoretically predicted that the vector potential of electrodynamics influences the interference
of electrons in regions, where the magnetic field is effectively zero.
For the history of experimental verification see monograph \cite{Tonomura Ahar book}. 

This shows that in quantum physics, in contrast to  classical physics, dynamics is determined by potentials themselves and not by their derivatives alone; this way the gauge structure gets physically significant.
\section{Large scale observations}

Large scale astronomy delivers new observational results, which are
relevant for general relativity as well as for quantum physics. Firstly,
the improved precise measurement of large scale distances by supernovae
Ia shows that the Hubble expansion is accelerated and thus a concrete value can be assigned to the time derivative of the Hubble function \( H(t)\) at our epoche.

Secondly, the determination of the masses of the galaxy superclusters
Great Attractor and Perseus Pisces leads to gravitational radii of
about thousand parsecs - the greatest values to handle so far. These
superclusters have a meaning for physics on Earth, which from a theoretical
viewpoint is quite fundamental. Their gravitational potentials are
constant in space through the entire solar system up to \( 10^{-11} \),
so they cannot yield observational effects in classical physics, because
their spatial gradients are practically zero. The absolute values
of these potentials, however, lie two or three orders of magnitude above that of the sun on Earth and well above the linear limit of general relativity.

The idea that the potential of the Great Attractor might lead to a gravitational Aharonov-Bohm effect has been developed in \cite{Ahluwgravcompl}. There the potential is added as an external one in the Schr{\"o}dinger equation. 

General relativity, however, is expressed mathematically in terms of differential geometry. Thus the question arises in which sense the nearly constant
potentials of the Perseus Pisces as well as the Great Attractor cluster  could be made accessible to differential geometry. The
answer lies in focussing on time dependence. One is accustomed to
observe the Hubble expansion through the recession of distant galaxies;
but there is a possibility to observe the Hubble flow locally on the
basis of the time dependence of exactly these gravitational potentials.
\section{Formulas and Values}

In more detail, let \( a(t) \) be the cosmological scale factor describing the average growth of physical distances. One expands \( a(t)\) around the present time \( t_{0} \), into a Taylor series \( a(t)=a(t_{0})+\dot{a}(t_{0})(t-t_{0})+\frac{1}{2}\ddot{a}(t_{0})(t-t_{0})^{2}+\ldots \). Using the definition of the Hubble function\begin{equation}
\label{Hubblefct}
H(t)=\frac{\dot{a}(t)}{a(t)}
\end{equation}
this gives up to second order 
\begin{equation}
\frac{a(t)}{a(t_{0})}=1+H_{0}(t-t_{0})-\frac{q_{0}}{2}H^{2}_{0}(t-t_{0})^{2} 
\end{equation}
with \( H_{0}=H(t_{0}) \) and \( q_{0}=-\frac{\ddot{a}(t_{0})}{a(t_{0})}\frac{1}{H_{0}^{2}} \) todays Hubble function and deceleration parameter respectively. From observation one obtains \( q_{0}=-\frac{1}{2} \), which, using (\ref{Hubblefct})
and the definition of \( q_{0} \), implies\begin{equation}
\label{diffHublHo}
\dot{H}(t_{0})=-\frac{1}{2}H_{0}^{2}.
\end{equation}
In the calculations below the best present estimate of \( H_{0} \) is chosen:
\begin{equation}
\label{valueh Hubble}
H_{0}=\frac{1}{1.3\times 10^{28}{\rm cm}}
\end{equation}
in geometric units.

The gravitational potential \( \phi _{GA} \) of the Great Attractor
supercluster in the solar system can be estimated to be \cite{KenyongaandK}\begin{equation}
\label{valGA}
\phi _{GA}=-\frac{r_{g}}{l}=-3.0\times 10^{-5},
\end{equation}
analysing data of streaming motions of galaxy clusters in \cite{Lyndon Bell}.
Hereby \( r_{g} \) is the gravitational radius in geometric units
and \( l \) the distance from the center of the Great Attractor region
to our Local Group.

Using mainly spectral methods, it is reported in \cite{Kraan Korteweg} that the Perseus Pisces cluster happens to lie approximately at similar
distances at opposite sides of the local group. 
\section{quantum physical curvature}

The Ansatz is the following. With \( z \) defining the redshift,
the relation \( z=H(t)\ l \) is used to gain the time dependence of the gravitational potentials \( \phi _{GA} \) and \( \phi _{PP} \) of the Great Attractor and the Perseus Pisces supercluster respectively. Now \( \phi _{GA} \)
becomes a function of \( t \),
\begin{equation}
\label{PhiGat}
\phi _{GA}(t)=-\frac{r_{g_{GA}}}{z}\ H(t),
\end{equation}
with \( \phi _{GA}(t_{0})=-3.0\times 10^{-5} \). For simplicity \( \phi _{PP} \)
is assumed to have the same value as \( \phi _{GA} \) and its time dependence is introduced in the same manner. Summing these two potentials and respecting Lorentzian signature a purely time dependent spacetime metric results
\begin{equation} 
g_{\mu\nu} = \left( \begin{array}{cccc} -
A(t)& 0& 0& 0 \\ 0& B(t)& 0& 0\\0&0&B(t)&0\\0& 0& 0&B(t)\end{array} \right)\label{metricdiag} 
\end{equation}
with \( A(t)=1+2\phi (t) \) and \( B(t)=1-2\phi (t) \) and the indices dropped.

 The component \( G^{00} \) of the Einstein tensor (\ref{Einsteintensor})  of this metric
can be calculated to be
\begin{equation}
\label{Go leading}
G^{00}=-\frac{3}{4}\frac{\dot{B}(t)^{2}}{A(t)^{2}B(t)^{2}}.
\end{equation}
It is a well known fact that the \( G^{00} \) and \( G^{0i} \) components of the Einstein tensor contain
only first derivatives of the metric with respect to time. Usually
the Einstein equations of these components are taken as constraint
equations. 

However, evaluation of \( G^{00} \) with (\ref{PhiGat}) and (\ref{diffHublHo}) at time \( t_{0} \) and constant redshift \(z\) gives in leading order
\begin{equation}
\label{Goeval}
G^{00}(t_{0})=-\frac{3}{4}\phi (t_{0})^{2}H_{0}^{2},
\end{equation}
then using (\ref{valueh Hubble}), (\ref{valGA}) and dividing by \( 8\pi  \) gives due
to Einstein field equations (\ref{Einsteinefeq}) 
\begin{equation}
\label{Too}
T^{00}=\frac{G^{00}}{8\pi }=-0.06\frac{2.6\times 10^{-66}}{{\rm cm}^{2}},
\end{equation}
which is nearly the value of the Casimir energy density per unit area and unit length given in (\ref{Casimirgu}).

Furthermore, \( G^{00} \) in (\ref{Go leading}) is due to the nonlinear
part of the Einstein tensor, which is usually neglected in linear
and Newtonian approximations of general relativity. But these approximations
imply flat background and thus run against the notion of intrinsically
curved spacetime.
\section{Conclusions and outlook}

Of course, this calculation is nonlocal in space, but 'locally' in
full four-dimensional spacetime.
One can insert the time dependent potentials in the post-Newtonian approximation developed in \cite{A&C38,EIH38}. Then nondiagonal terms would reflect the deviation of Perseus Pisces and Great Attractor being placed on a line. Nevertheless, \( G^{00} \), but not \( G_{00} \), would remain to
be the same.

Also this Ansatz can be used to investigate differential topological properties of spacetime. This will be discussed in a
future publication \cite{thesis}.

As a result we conclude, that Planck's constant appears to emerge
from spacetime and represents the obstruction to flatness which is
characteristic for the notion of curvature

Furthermore, as reported in \cite{woudtdebates} there are ambiguities regarding extent and nature of the Great Attractor which are lively discussed among astronomers. Also an influence of the Perseus Pisces cluster on the Local Group may be questionable, as stated for example in \cite{tonrysbftwo}. Perhaps this metric may be helpful to clarify the structure of the galaxy superclusters as well as the motion of the Local Group.

\section{Acknowledgment}
This work has been supported by the Hans B{\"o}ckler foundation.

\bibliography{apssamp}

\end{document}